\documentclass{article}
\usepackage{spconf,amssymb,amsmath,epsfig}
\usepackage{textcomp}
\usepackage{amsfonts,amsthm,amsopn}
\usepackage{bm}
\usepackage{hyperref}
\usepackage{url}
\usepackage[capitalise]{cleveref}
\usepackage{graphicx,caption,lipsum}
\usepackage{booktabs,multirow}
\usepackage{placeins}
\usepackage{algpseudocode,algorithm}
\usepackage{color}
\usepackage[table,xcdraw]{xcolor}
\usepackage{bbm}
\usepackage{enumerate}   
\usepackage{adjustbox}
\usepackage[normalem]{ulem}
\usepackage{fixltx2e}
\usepackage[toc,page]{appendix}
\usepackage{makecell}
\usepackage{soul}

\usepackage{color}
\definecolor{mypink1}{rgb}{0.858, 0.188, 0.478}

\title{Improving Transformer-Based Networks with Locality for Automatic Speaker Verification}

\name{{Mufan Sang$^{1\ast}$, Yong Zhao$^{2}$, Gang Liu$^{2}$, John H.L. Hansen$^{1}$, Jian Wu$^{2}$}}
\address{$^{1}$The University of Texas at Dallas, TX, USA \\
$^{2}$Microsoft Corporation, One Microsoft Way, Redmond, WA 98052, USA \\
{\small \tt \{mufan.sang,john.hansen\}@utdallas.edu},
{\small \tt \{yonzhao,ganli,jianwu\}@microsoft.com}}

\newcommand{\mfmod}[1]{{\color{red}{\emph{#1}}}}
\renewcommand{\mfmod}[1]{}

\graphicspath{{./image/}}

\begin{document}

\maketitle

\ninept

\begin{abstract}
Recently, Transformer-based architectures have been explored for speaker embedding extraction. Although the Transformer employs the self-attention mechanism to efficiently model the global interaction between token embeddings, it is inadequate for capturing short-range local context, which is essential for the accurate extraction of speaker information. In this study, we enhance the Transformer with the enhanced locality modeling in two directions. First, we propose the Locality-Enhanced Conformer (LE-Confomer) by introducing depth-wise convolution and channel-wise attention into the Conformer blocks. Second, we present the Speaker Swin Transformer (SST) by adapting the Swin Transformer, originally proposed for vision tasks, into speaker embedding network. We evaluate the proposed approaches on the VoxCeleb datasets and a large-scale Microsoft internal multilingual (MS-internal) dataset. The proposed models achieve 0.75\% EER on VoxCeleb 1 test set, outperforming the previously proposed Transformer-based models and CNN-based models, such as ResNet34 and ECAPA-TDNN. When trained on the MS-internal dataset, the proposed models achieve promising results with 14.6\% relative reduction in EER over the Res2Net50 model.
\end{abstract}

\begin{keywords}
Speaker verification, self-attention, Transformer, Conformer, Swin Transformer
\end{keywords}
%

\section{Introduction}
In the past few years, we have seen the fast development of deep neural network (DNN) based speaker verification (SV) systems. Various models for speaker verification have been proposed with different DNN architectures~\cite{snyder2018x, zeinali2019but, desplanques2020ecapa}, novel loss functions~\cite{wan2018generalized, chung2020defence, wang2018cosface}, and frameworks for overcoming domain mismatch~\cite{sang2020open, bhattacharya2019generative, sang2021deaan}. In~\cite{inoue2020semi, sang2022self, chen22g_interspeech}, researchers further studied semi-supervised and self-supervised SV systems using partially labeled or unlabeled data.

\mfmod{CNN-based architectures have been the predominant model for SV system, such as x-vector, ResNet-based architectures, and ECAPA-TDNN }  

\renewcommand{\thefootnote}{\fnsymbol{footnote}}
\footnotetext[1]{Work performed while Mufan Sang was an intern at Microsoft.}

\mfmod{Generally, an entire speaker verification system consists of a front-end speaker embedding network to extract discriminative speaker embeddings and a back-end scoring module to calculate the similarity or measure a score between embedding pairs. }For speaker embedding extraction, x-vector~\cite{snyder2018x} has been proven successful by utilizing the 1D convolutional neural network (CNN) based time delay neural network (TDNN). Moreover, 2D CNNs (i.e. ResNet-based architectures) are also successfully adopted to the SV task and obtain remarkable performance~\cite{cai2018exploring, chung2020defence}. ECAPA-TDNN~\cite{desplanques2020ecapa} was proposed to further enhance the TDNN-based architecture and achieved a competitive performance with ResNet. These predominant CNN-based models take advantage of strong ability of 
capturing local speaker patterns from speech features. To further improve the performance of CNN-based speaker embedding networks, some attention mechanisms were integrated to the speaker embedding extractor~\cite{zhou2019deep, sang22_interspeech} or the pooling layer~\cite{cai2018exploring,okabe2018attentive}. Convolution layer allows CNNs to model the local dependencies well, but it lacks a mechanism to capture speaker information globally\mfmod{capture long-range global context}. There have been attempts to explore using Transformer to replace CNNs for speaker embedding extraction~\cite{mary2021s, safari2020self, wang2022multi}. However, without large-scale pre-training, Transformer-based speaker embedding networks can hardly achieve competitive performance as CNNs for speaker verification. This is primarily due to the lack of certain desirable properties inherently built into the CNN architecture such as locality. 

In this study, we aim to enhance the Transformers in capturing global and local context collectively. We introduce locality mechanisms to Transformer in two directions. First, we propose the Locality-Enhanced Conformer (LE-Confomer) by incorporating depth-wise convolution and channel-wise attention into the feed-forward network (FFN) of the  Conformer block~\cite{gulati2020conformer}. We investigate an effective way to aggregate the output features from all the LE-Conformer blocks to improve the frame-level speaker embedding. Second, we present Speaker Swin Transformer (SST) by employing the hierarchical Transformer architecture with shifted local window self-attention, inspired by Swin Transformer~\cite{liu2021swin}. 

Experimental results on the VoxCeleb datasets demonstrate that the proposed LE-Conformer and SST significantly outperform the previously proposed Transformer-based models and ResNet and ECAPA-TDNN baseline systems. Moreover, when trained on a larger-scale MS-internal multilingual dataset, the proposed systems outperform Res2Net50 by a large margin, producing more robust and competitive speaker embeddings. The primary contributions of this paper can be summarized as follow: (1) We propose an effective locality mechanism for Conformer to enhance the ability of local information aggregation. (2) We propose the Speaker Swin Transformer which generates multi-scale output feature maps with shifted local window self-attention. (3) We conduct comprehensive experiments to demonstrate the effectiveness of the proposed Transformer-based networks with locality mechanisms. 


\begin{figure*}[th]
\centering
\scalebox{0.95}
{
\includegraphics[width=18.0cm,height=6.6cm]{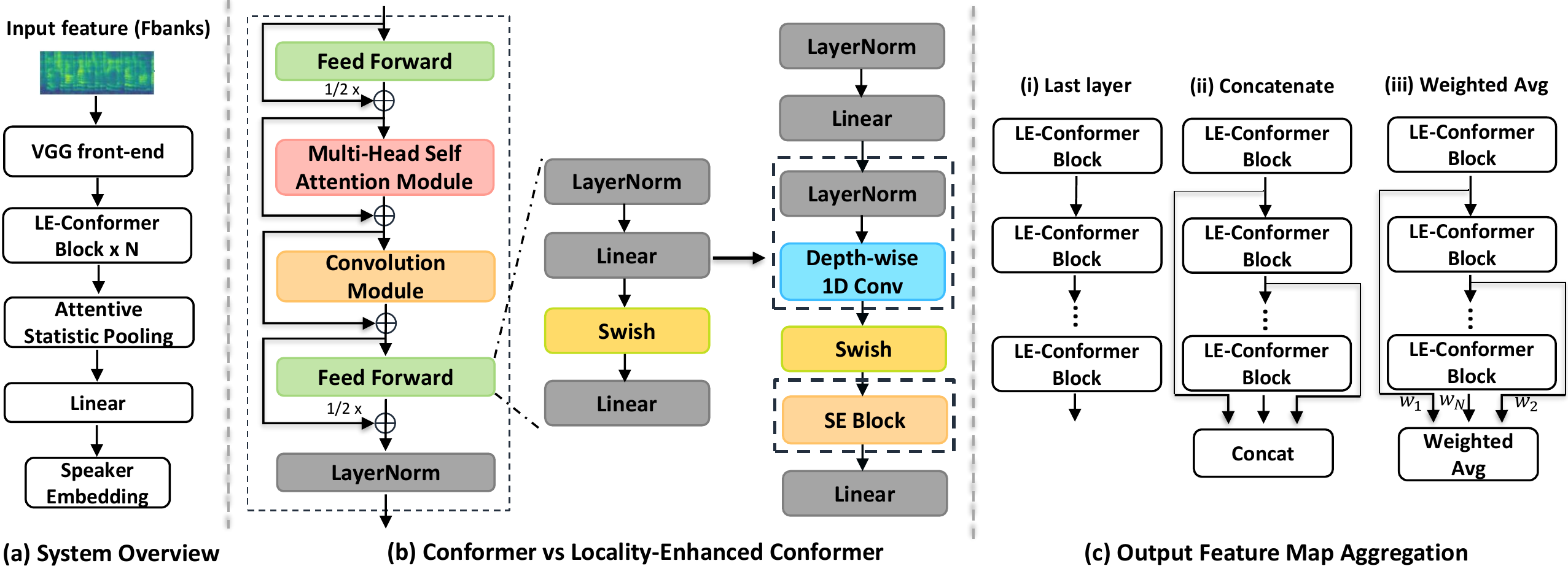}
}
\vspace{-2.0mm}
\caption{The architecture of Locality-Enhanced Conformer for speaker embedding.} 
\label{fig:system}
\end{figure*}
\vspace{-0.5ex}

\section{Locality-Enhanced Conformer}
Self-attention is the key component of the Transformer. It enables the Transformer to have a strong ability to model the global interaction between speech frames. However, global self-attention does not have sufficient ability to capture local information which is essential for speaker embedding. Therefore, we introduce locality mechanisms to enhance Transformer in modeling local dependencies. In this section, we present Locality-Enhanced Conformer (LE-Confomer), which is built upon the Conformer~\cite{gulati2020conformer} architecture. We incorporate additional convolution, channel-wise attention, and intermediate output feature map aggregation into the Conformer network. The system architecture is illustrated in Fig. \ref{fig:system}(a). It comprises (i) a VGG front-end to subsample the input feature, (ii) a number of locality-enhanced Conformer blocks to extract frame-level speaker embedding, (iii) a pooling layer to generate utterance-level embedding, (iv) a linear layer to extract the final speaker embedding. 



\subsection{Conformer Encoder Block}
As a state-of-the-art model in ASR, Conformer~\cite{gulati2020conformer} combines convolution and self-attention in the Transformer block to enhance its capability of capturing local information. The architecture of Conformer block is shown in the left side of Fig. \ref{fig:system}(b). It consists of multi-head self-attention (MSA) module and convolution module sandwiching by two Macron-style feed-forward networks (FFN). Assuming $\boldsymbol{z_{i}}$ as the input for the $i$-th Conformer block, the output of this block $\boldsymbol{z_{i+1}}$ is computed as

\begin{equation}
\begin{aligned}
\tilde{\boldsymbol{z}_i} &=\boldsymbol{z_i}+\frac{1}{2} \mathrm{FFN}\left(\boldsymbol{z_i}\right) \\
\boldsymbol{z_i}^{\prime} &=\tilde{\boldsymbol{z_i}}+\operatorname{MSA}\left(\tilde{\boldsymbol{z_i}}\right) \\
\boldsymbol{z_i}^{\prime \prime} &=\boldsymbol{z_i}^{\prime}+\operatorname{Conv}\left(\boldsymbol{z_i}^{\prime}\right) \\
\boldsymbol{z_{i+1}} &=\operatorname{LayerNorm}\left(\boldsymbol{z_i}^{\prime \prime}+\frac{1}{2} \mathrm{FFN}\left(\boldsymbol{z_i}^{\prime \prime}\right)\right)
\end{aligned}
\end{equation}
where FFN denotes the feed-forward network, MSA denotes the multi-head self-attention, and Conv denotes the convolution module.  

\subsection{Introducing Locality and Channel-wise Attention}
In each block, adding convolutional layers after MSA sequentially helps the Conformer capture local dependencies. Regarding the FFN module, hidden dimension of the latent feature is expanded between the two linear layers. We consider that depth-wise convolution could be a filter for the latent representation and introduce additional locality to Transformer. To achieve this goal, we propose an effective strategy to integrate depth-wise convolution and a channel-wise attention mechanism into the feed-forward network. 

As shown in the right side of Fig. \ref{fig:system}(b), LayerNorm and a 1D depth-wise convolution layer are added after the first linear layer of FFN. The depth-wise convolution provides information interaction among adjacent frames, and captures local continuity of input feature maps. Integrating Squeeze-and-Excitation (SE) block~\cite{hu2018squeeze} after convolution can further improve representation ability by modeling the inter-dependencies between channels of feature maps and re-calibrating each channel.

\subsection{Aggregating Output Feature Maps of LE-Conformer Blocks}
For the frame-level speaker embedding extractor, the output of the last layer is often used as the input for the pooling layer. However, previous studies~\cite{gao2019improving, tang2019deep} indicated that feature maps from lower layers can also contribute to robust speaker embeddings. Similar to ~\cite{zhang2022mfa}, we apply two strategies to utilize low-layer feature maps. As illustrated in Fig. \ref{fig:system}(c), the first strategy is to concatenate the output feature maps from all LE-Conformer blocks along the channel dimension. Secondly, we conduct weighted average of the output feature maps from all LE-Conformer blocks with learnable weights. 


Finally, the enhanced frame-level speaker embedding is processed by the pooling layer to generate utterance-level speaker embedding.   

\begin{figure}[t]
\centering
\scalebox{1.0}
{
\includegraphics[width=8.6cm,height=8.0cm]{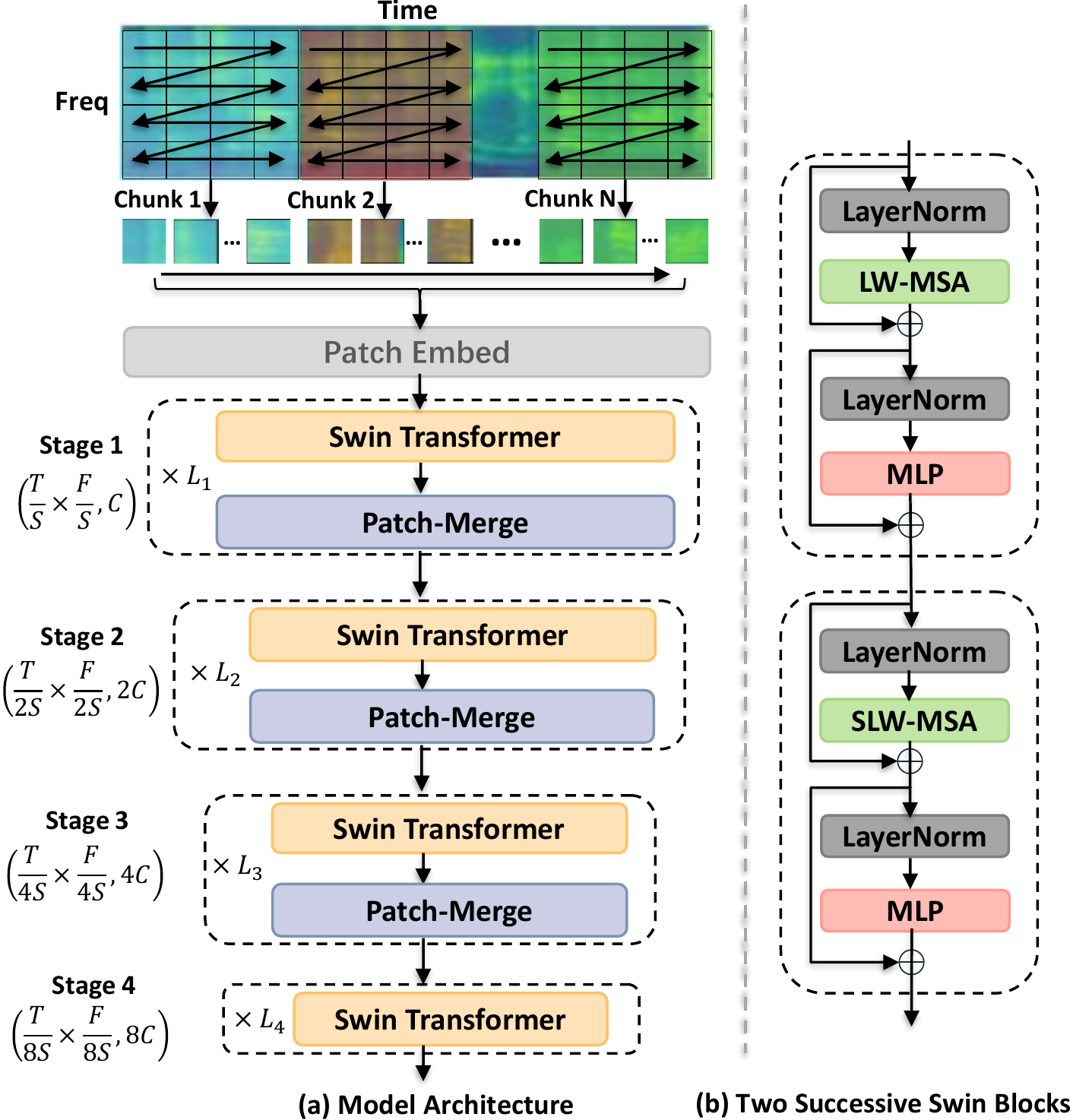}
}
\vspace{-6.0mm}
\caption{The architecture of Speaker Swin Transformer.} 
\label{fig:Swin}
\end{figure}

\vspace{-2.0ex}
\section{Speaker Swin Transformer}
\vspace{-1.0ex}
Generally, the Transformer computes the self-attention globally which contributes to a great long-range dependencies modeling capability. However, speech features can be much longer than text sentences. Thus, Transformers usually incur high computational and memory costs for speech tasks. Moreover, with convolution operation, ResNet-based networks bring locality, inductive bias, and multi-scale feature maps which contribute to their successes on SV tasks. On the contrary, Transformer's output feature maps are single-scale among all blocks. With a hierarchical structure and shifted window self-attention, Swin Transformer~\cite{liu2021swin} has achieved state-of-the-art performance on multiple vision tasks. Therefore, we bring the advantages of Swin Transformer to speaker verification, proposing the Speaker Swin Transformer (SST), which generates multi-scale output features and computes self-attention with shifted local windows. The architecture leads to linear computational complexity to the length of utterance. The overview of the Speaker Swin Transformer encoder is presented in Fig. \ref{fig:Swin}.

\subsection{Encoding Input Features by Overlapping Patch Embedding}
Different from Conformer, Speaker Swin Transformer processes input speech features patch-by-patch instead of frame-by-frame. \mfmod{Swin Transformer splits the input image into non-overlapping patches.}Considering the characteristics of speech feature, using patch embedding as input token allows model to learn both the temporal and frequency structure. As shown in Fig. \ref{fig:Swin}(a), to better capture local continuity of input feature, the input feature is split into a sequence of $P\times P$ ($P=7$) patches with overlap by half of its area. Each patch is flattened and projected into a $C$-dimensional ($C=96$) 1D patch embedding with a linear patch embedding layer. We utilize a convolution layer with $7\times 7$ kernel size, a stride of 4 ($S=4$), and zero padding of 3 for patch embedding. Accordingly, the output size of patch embedding is $\left(\frac{T}{S} \times \frac{F}{S}, C\right)$, where $T$ and $F$ represent time and frequency domain of input feature. Although Fbanks and images have a similar format with 2D shape, the height and width of Fbanks contain different information which represents frequency and temporal dimension. Inspired by~\cite{chen2022hts}, to better model the dependencies among frequencies for nearby frames, we first split the whole Fbanks into chunks along temporal dimension. Then, patches are split within each chunk following the order shown in Fig. \ref{fig:Swin}(a), and all patches compose the sequence chunk-by-chunk. Then, the patch token sequence is processed by the linear patch embedding layer and sent to the following Transformer blocks.          


\subsection{Swin Transformer Block}
In order to construct the hierarchical structure with local window self-attention, we adapt Swin Transformer block to speaker verification. Inside each Swin Transformer block, the local window self-attention is introduced to replace the conventional global self-attention for efficient modeling. Regarding self-attention computing, all the input patch tokens are evenly partitioned into non-overlapping local windows which contain $M\times M$ ($M=5$) patches. Accordingly, self-attention is computed within each local window instead of among all patches globally. For an input speech feature with $f\times t$ patch tokens, the computational complexity of a global MSA is $\mathcal{O}\left(f t C^2+(f t)^2 C\right)$ and a local window MSA (LW-MSA) is $\mathcal{O}\left(f t C^2+M^2 f t C\right)$ with the latent feature dimension $C$. It demonstrates that LW-MSA is much more efficient than global MSA with $M^2 \ll ft $ ($M=5$) and its complexity grows linearly with $ft$ instead of quadratically as global MSA. Therefore, it is able to substantially decrease the computation cost.  

To enlarge the receptive field and model the connections across windows, shifted local window multi-head self-attention (SLW-MSA) is introduced in addition to the LW-MSA. This module adopts a windowing configuration that is shifted from that of the preceding layer, by shifted toward lower-right by $\left(\left\lfloor\frac{M}{2}\right\rfloor,\left\lfloor\frac{M}{2}\right\rfloor\right)$ patches in consecutive Swin Transformer blocks. As shown in Fig. \ref{fig:Swin}(b), the Transformer block with SLW-MSA introduces connections between neighboring non-overlapping windows from the previous block. 

To form a hierarchical structure, a patch merging layer~\cite{liu2021swin} is added at the end of each stage from stage 1 to 3. For the first patch merging layer, it concatenates the neighboring patches with group of $2 \times 2$, and applies a linear layer to reduce the output dimension from $\left(\frac{T}{2S} \times \frac{F}{2S}, 4C\right)$ to $\left(\frac{T}{2S} \times \frac{F}{2S}, 2 C\right)$. As illustrated in Fig. \ref{fig:Swin}(a), the shape of patch tokens is reduced to $\left(\frac{T}{4S} \times \frac{F}{4S}, 4C\right)$ and $\left(\frac{T}{8S} \times \frac{F}{8S}, 8C\right)$ through stage 2 and 3, respectively. These stages allow the network to generate hierarchical representations as ResNet, so the memory cost is decreased exponentially through each stage. More importantly, the receptive field for local window self-attention grows larger as layer goes deeper. In summary, these designs contribute to efficient local and global relations modeling. 




\section{Experiments}
\vspace{-1.5ex}
\subsection{Datasets}
\textbf{\textit{VoxCeleb}} The SV systems are trained on the development set of VoxCeleb1\&2~\cite{nagrani2017voxceleb, chung2018voxceleb2} and evaluated on VoxCeleb1 test set. The total duration of the training data is around 2k hrs. We augment the training data by distorting the raw speech with additive noise~\cite{snyder2015musan}, RIR reverberation~\cite{ko2017study} and speed perturbation (with speed factor 0.9 and 1.1). Specifically, the generated utterances by speed perturbation are regarded as from new speakers. 

\noindent{\textbf{\textit{MS-internal}} This is a large-scale Microsoft internal multilingual dataset collected in controlled acoustic environments. It consists of around 26 million utterances from over 29k speakers in 48 languages and locales, about 4.6 seconds in length per utterance. The training set contains around 24 million utterances from over 27k speakers with total duration of 33k hrs, and the test set contains over 0.1 million utterances from around 1.7k speakers. There is no speaker overlapping between training and testing set.}

\subsection{Implementation Details}
The input features are 80-dimensional log Mel-filterbanks with a frame-length of 25 ms and 10 ms shift, applied with mean normalization at the utterance level. The proposed Transformers serve as the backbone network. Attentive statistic pooling  (ASP)~\cite{okabe2018attentive} is used to generate utterance-level embeddings. The models are trained with additive margin softmax (AM-softmax) loss~\cite{wang2018cosface} with a margin of 0.2 and a scaling factor of 30. We also trained  ResNet34, ECAPA-TDNN, and Res2Net50 for comparison.   


\mfmod{For the Conformer-based speaker embedding extractor,}
\textbf{\textit{Locality-Enhanced Conformer}} A segment of 2.0 seconds is randomly selected for each input utterance. The model consists of 6 Locality-Enhanced Conformer encoder blocks with 4 attention heads. For each block, we set the encoder dimension as 512, the kernel size of convolution module as 15, and the hidden unit size as 2048 for the FFN. We use the AdamW~\cite{loshchilov2018decoupled} optimizer with an initial learning rate of 3e-4 and set the weight decay as 5e-2. A linear warm-up is applied at the first 45k steps and the learning rate is adjusted based on cyclical annealing schedule in the range of 1e-8 and 3e-4. The batch size is 128 for each of 8 GPU cards.

\textbf{\textit{Speaker Swin Transformer}} A segment of 3.2 seconds is randomly selected for each input utterance. The input feature with size $320\times 80$ will be equally split into 2 chunks along time dimension with size $160\times 80$. We set patch size as $7\times 7$, \mfmod{To adapt the patch chunk with size $160\times 80$ to the hierarchical structure, }and attention window size as $5\times 5$. The four network stages are designed with 2, 2, 6, 2 Swin Transformer blocks, respectively. We set the channel number of the hidden layer in the first stage $C=96$. We utilize AdamW optimizer with an initial learning rate of 3e-2 and set the weight decay as 5e-2. A linear warm-up is applied at the first 130k steps, and the learning rate is adjusted based on cyclical annealing schedule in the range of 1e-5 and 3e-2. The batch size is 200 for each of 8 GPU cards. 

We report the system performance using two evaluation metrics: Equal Error Rate (EER) and minimum Detection Cost Function (minDCF) with $p_{target}=0.05$. \mfmod{Cosine similarity is adopted for scoring in the testing phase.}

\vspace{-1.0ex}
\subsection{Evaluation on VoxCeleb Dataset}
\vspace{-1.0ex}


\begin{table}[t]
\caption{Performance of all SV systems on VoxCeleb1. \textit{Upper block: CNN-based models; Middle block: Transformer-based models; Lower block: Our proposed models}}
\setlength{\tabcolsep}{3.2mm}{
\renewcommand\arraystretch{1.2}
\scalebox{0.97}{
\begin{tabular}{lccc}
\hline \textbf{Systems} & \textbf{Corpus} & \textbf{EER(\%)} & \textbf{minDCF} \\
\hline 
ResNet34~\cite{zeinali2019but} & Vox1\&2 & $1.06$ & $0.084$ \\
ECAPA-TDNN~\cite{desplanques2020ecapa} & Vox1\&2 & $0.85$ & $0.078$ \\
Res2Net50 {~\cite{zhou2021resnext}} & Vox1\&2 & $0.55$ & $0.041$ \\
\hline
SAEP {~\cite{safari2020self}} & Vox2 & $5.44$ & $-$ \\
Wang et al.{~\cite{wang2022multi}} & Vox2 & $2.56$ & $-$ \\
DT-SV {~\cite{zhang2022dt}} & Vox2 & $1.92$ & $0.130$ \\
S-vector+PLDA {~\cite{mary2021s}} & Vox1\&2 & $2.67$ & $0.300$ \\
\hline
LE-Conformer (ours) & Vox1\&2 & $\mathbf{0.75}$ & $\mathbf{0.055}$ \\
SST (ours) & Vox1\&2 & $1.34$ & $0.104$ \\
\hline
\end{tabular}}
}
\label{table:Vox}
\end{table}

First, we evaluate the proposed LE-Conformer and SST on the  VoxCeleb dataset. Table \ref{table:Vox} compares the performance of our systems with CNN-based baseline systems ResNet34~\cite{zeinali2019but}, ECAPA-TDNN~\cite{desplanques2020ecapa}, Res2Net50~\cite{zhou2021resnext} and other Transformer models~\cite{mary2021s, safari2020self, wang2022multi, zhang2022dt} for speaker verification. The LE-Conformer and SST achieve 0.75\% and 1.34\% EER, respectively, which significantly outperform other Transformer-based systems. Compared to S-vector+PLDA, LE-Conformer and SST improve the performance with relative 71.9\% and 49.8\% reduction in EER. Moreover, the LE-Conformer and SST also outperform the Wang et al~\cite{wang2022multi} with relative 70.7\% and 47.7\%, and DT-SV with 60.9\% and 30.2\% improvement in EER. 

Compared to CNN-based networks, LE-Conformer outperforms ResNet34 and ECAPA-TDNN with relative 29.3\% and 11.8\% improvement in EER, respectively. It demonstrates that introducing locality mechanisms to the Transformer is beneficial for more accurate extraction of speaker embeddings.

\begin{table}[h]
\caption{Ablation study of the Locality-Enhanced Conformer and Speaker Swin Transformer. \textit{Non-OPE: non-overlapping patch embedding.}}
\vspace{-4.0mm}
\setlength{\tabcolsep}{3.5mm}{
\begin{center}
\renewcommand\arraystretch{1.05}
\scalebox{0.99}{
\begin{tabular}{lcc}
\hline \textbf{Systems} & \textbf{EER(\%)} & \textbf{minDCF} \\
\hline LE-Conformer & $0.75$ & $0.055$ \\
\hspace{0.5em} No SE Block & ${0.87}$ & ${0.060}$ \\
\hspace{0.5em} No DW Conv & ${0.94}$ & ${0.068}$ \\
\hspace{0.5em} No Concat & ${1.00}$ & ${0.070}$ \\
\hspace{0.5em} Weighted avg & ${1.21}$ & ${0.091}$ \\
\hline SST  & $1.34$ & $0.104$ \\
\hspace{0.5em} Non-OPE & $1.47$ & $0.120$ \\
\hline
\end{tabular}}
\end{center}}
\vspace{-3mm}
\label{table:Ablation}
\end{table}

\begin{table}[h]
\caption{Performance of the proposed systems on MS-internal dataset.}
\vspace{-4.0mm}
\setlength{\tabcolsep}{4mm}{
\begin{center}
\renewcommand\arraystretch{1.0}

\begin{tabular}{l|cc}
\hline \textbf{Systems} & \textbf{EER} (\%) & \textbf{minDCF} \\
\hline Res2Net50 & $3.09$ & $0.180$ \\
LE-Conformer & $3.57$ & $0.229$ \\
SST & $\mathbf{2.64}$ & $\mathbf{0.168}$ \\

\hline
\end{tabular}
\end{center}}

\label{table:KingASR}
\end{table}

We further conduct experiments to illustrate the effectiveness of the proposed speaker embedding networks, and how the local information helps to improve the performance of Transformer-based models. Table \ref{table:Ablation} shows the impact of changes to the LE-Conformer block and SST. For LE-Conformer, without adding SE block and depth-wise convolution into the FFN, the performance degrades with an increase of 25.3\% and 23.6\% relatively in EER and minDCF, respectively. \mfmod{We also study the effective way to aggregate outputs from Conformer blocks.}Without concatenating the output feature maps from all blocks, the performance further degrades by 6.4\% relative in EER. It is equivalent to Conformer after removing the SE block, DW conv, and concatenation. The results in Table \ref{table:Ablation} demonstrate that the performance of Transformer can be significantly improved by the integration of locality mechanisms. Moreover, the performance becomes even worse if output features are averaged with learnable weights before the pooling layer.

For SST, it is beneficial to use overlapping patch embedding (OPE), which yields 8.8\% relative improvement in EER compared to non-overlapping patch embedding (non-OPE). The overlapping patch embedding can effectively enhance the capability of modeling local continuity of input features via overlapped sliding windows.  






\vspace{-1.0ex}
\subsection{Evaluation on MS-internal Dataset}
\vspace{-1.0ex}
In this section, we investigate the performance of proposed models when trained on a large-scale Microsoft internal multilingual (MS-internal) dataset. We compare LE-Conformer and SST to Res2Net50, the best CNN-based system in Table \ref{table:Vox}. As illustrated in Table \ref{table:KingASR}, Speaker Swin Transformer outperforms Res2Net50 by 14.6\% relative improvement in EER. It demonstrates that the hierarchical transformer structure with shifted local window self-attention is capable of making full use of the massive training data and learning global and local information collectively, compared with CNN-based models.

\mfmod{the performance of these systems trained on the MS-internal dataset.}

\vspace{-2.0ex}
\section{Conclusions}
\vspace{-1.5ex}
In this paper, we propose two speaker embedding networks by incorporating locality mechanisms to Transformer. Firstly, we integrate depth-wise convolution and channel-wise attention into the Conformer blocks to enhance the ability of modeling local dependencies. Secondly, we introduce the Speaker Swin Transformer which processes input features at multiple scales with shifted local window self-attention. Experimental results demonstrate that our models significantly outperform previous Transformer-based models and CNN-based models, such as ResNet34 and ECAPA-TDNN. The proposed architectures enable the effective extraction of speaker embeddings, especially when trained on large amounts of data. We hope this work can provide inspirations for the ultimate design of Transformer-based speaker embedding networks.

\vfill\pagebreak

\footnotesize
\bibliographystyle{IEEEbib}
\bibliography{icassp21}


\end{document}